\documentclass[10pt,notitlepage,showpacs,pra,twocolumn]{revtex4-1}
\usepackage{ulem}
\usepackage{amsfonts}
\usepackage{amsmath}
\usepackage{amssymb}
\usepackage{graphicx}
\usepackage{float}
\usepackage[toc,page,header]{appendix}

\begin{document}

\title{Robust switching of superposition-states via a coherent double stimulated Raman adiabatic passage}
\author{Luyao Yan, Dandan Ma, Dongmin Yu, Jing Qian$^{\dagger}$ }
\affiliation{Department of Physics, School of Physics and Material Science, East China
Normal University, Shanghai 200062, People's Republic of China}

\begin{abstract}
Coherent manipulation of quantum states is of crucial importance in accurate control of a quantum system. A fundamental goal is coherently transferring the population of a desired state with near-unit fidelity. For this propose, we theoretically demonstrate a novel coherent double-stimulated Raman adiabatic passage (STIRAP) in a three-level $\Lambda$-type system for realizing the switch of unequal initial preparations on two ground states. This operation uses single optical pulse sequence accomplishing both bright-STIRAP and dark-STIRAP process, in which the intermediate-level detuning and the pulse delay must be given an optimal adjustment. Besides, owing to the imperfection of double-STIRAP transition, the sensitivity of the switch fidelity with respect to the spontaneous loss from the intermediate state, to the pulse amplitude and to the population difference, are also discussed. This work suggests a simple and experimentally-feasible all-optical approach to switch the superposition quantum state, serving as one-step closer to the goal of coherent manipulation of quantum systems.
\end{abstract}
\email{jqian1982@gmail.com}
\maketitle
\preprint{}

\section{Introduction}

With the growing interests in quantum information, preparation and manipulation of specified coherent superposition of quantum states have become an important task. A particularly robust and powerful technique is dark-stimulated Raman adiabatic passage (d-STIRAP) \cite{Bergmann98,Vitanov17}, by which the population of a target state can be obtained with near-unit efficiency via a long-time adiabatic evolution along a dark state. An ordinary d-STIRAP enables a complete coherent population transfer from one state to another irrelevant to the radiative loss from other middle states, which can be sped up with a shortcut-to-adiabatic protocol via shaped pulses \cite{Du16}. Accounting for its novel properties, d-STIRAP and its extensions to multi-level systems are treated as the basic tools for coherent state manipulation, providing a great number of applications in the field of quantum information science, such as producing quantum gate with Rydberg atoms \cite{Beterov13,Goerz14} and the entangled-state preparation \cite{Marr03,Chen07,Moller08b, Zhao17}. A specific example is properly controlling the accumulated geometric phase in d-STIRAP can efficiently change the quantum states, presenting robust geometric phase gates for quantum computing \cite{Moller07,Moller08a} and solid-state systems \cite{Yale16,Coto17}.

Besides the d-STIRAP, which is immune against loss through spontaneous emission from the intermediate state, an alternative approach as so-called bright-STIRAP (or b-STIRAP) for population transfer simultaneously arises, which relies on an intuitive pulse sequence \cite{Klein07,Grigoryan09,Chakhmakhchayn12}. This b-STIRAP is a faster process which features a sufficiently large intermediate-level detuning and sufficiently short transition time to suppress the effect of intermediate state, the importance of which are also discussed in refs. \cite{Vitanov97,Boradjiev10,Militello11}.

As for a single quantum system implemented by both d-STIRAP and b-STIRAP, most previous works adopt double adiabatic sequences where the first d-STIRAP is used to deterministically prepare the target state and the second reversed b-STIRAP pulse is used to bring the system back to the initial state for measurement after a wait time. Such a protocol is popular in generating ultracold molecules \cite{Winkler07,Ospelkaus08,Danzl08}. Moreover, by varying the wait time, a single-qubit geometric phase gate can be created by coherent excitation of a trapped Rydberg ion \cite{Higgins17}, even yielding complex NOON state entanglement between two Bose-Einstein condensates \cite{Idlas16}. 
\textbf{On the other hand, Du and coworkers experimentally implemented a double-STIRAP process realized by a single laser pulse sequence in a three-level $\Lambda$ system with same transfer efficiency for both d- and b-STIRAPs, suggesting the possibility for coherent manipulation of quantum superposition states with high efficiency \cite{Du14}. However, a detailed discussion for the essence of the whole switching process and its relevance to system parameters have not been presented}

In the present work, after an optimal control for relevant parameters, we represent a robust and simple way to change the population of two components of an initial superposition state in a three-level $\Lambda$ configuration by utilizing a novel double-STIRAP. \textbf{Different from the previous schemes such as double detuning-induced STIRAP where time-dependent detuning pulses are used to preserve the coherence in hyperfine levels \cite{Deng17}, or a double-pass approach where a second reversed pulse sequence is used for measuring the single-pass transition probability \cite{Vitanov18}}, we apply a single pulse sequence to implement the b-STIRAP and d-STIRAP at the same time, in order to efficiently switch an arbitrary superposition state in a way of $\left\vert\Phi_{t=0}\right\rangle = \alpha\left\vert a\right\rangle + \beta\left\vert b\right\rangle$ $\to$ $\left\vert\Phi_{t=+\infty}\right\rangle = e^{i\delta\phi}\beta\left\vert a\right\rangle + \alpha \left\vert b\right\rangle$. The intermediate-level detuning and the pulse sequence delay must satisfy the conditions of adiabaticity and b-STIRAP transition, and the sensitivity of the scheme to other parameters such as the spontaneous loss of intermediate state, the laser pulse amplitudes, is also illustrated. Finally, we show a practical and robust operation to a fast switching of population of the two components of the state with a time period of $4.5\mu$s and a high-fidelity $>0.99$, indicating its wider perspectives to the coherent manipulation of quantum systems.

\section{Scheme description} 

\subsection{Eigenstates and double-STIRAP}

\begin{figure}
\includegraphics[width=3.4in,height=3.3in]{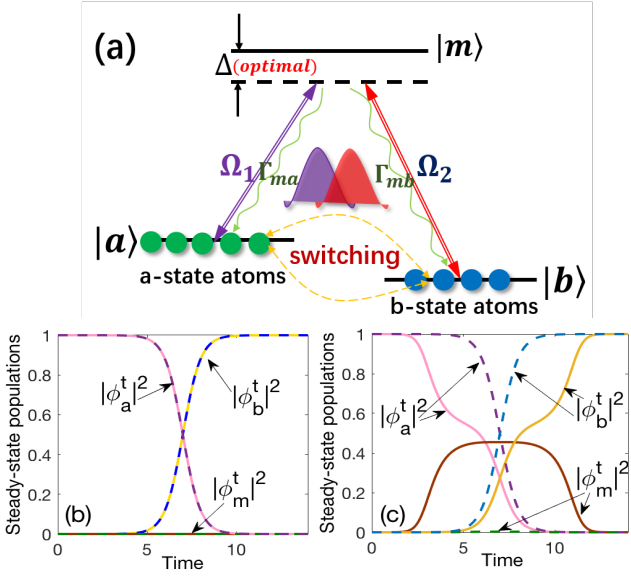}
\caption{(color online). (a) Schematic representation of a three-level $\Lambda$-type configuration with relevant energy levels and optical couplings. Relevant parameters are described in the text. (b) Steady-state population $|\phi_i^t|^2$ of state $\left\vert i\right\rangle$ in d-STIRAP.  $\Delta=0.1\Omega$ (small-detuning) and $\Delta = 10\Omega$ (large-detuning) are denoted by solid and dashed lines, respectively. (c) is same to (b) except for describing the case of b-STIRAP.}
\label{modelfig}
\end{figure}

The underlying physical mechanism is illustrated by Fig. \ref{modelfig}(a), \textbf{where we consider a three-level $\Lambda$-type model with two ground states $\left\vert a\right\rangle$, $\left\vert b\right\rangle$ coupled to a common excited state $\left\vert m\right\rangle$ by a pair of lasers with Rabi frequencies $\Omega_{1}$ and $\Omega_{2}$, with respect to the transitions of $\left\vert a\right\rangle\leftrightarrow\left\vert m\right\rangle$ and $\left\vert b\right\rangle\leftrightarrow\left\vert m\right\rangle$.} Under the prerequisite of the rotating-wave approximation and two-photon resonance, the Hamiltonian is given by ($\hbar=1$) 
\begin{equation}
\hat{H}=\Delta\hat{\psi}_m^\dagger\hat{\psi}_m+\frac{\Omega_{1}}{2}(\hat{\psi}_m^\dagger\hat{\psi}_a e^{i\phi_L}+h.c.)+\frac{\Omega_{2}}{2}(\hat{\psi}_b^\dagger\hat{\psi}_m+h.c.) \label{Ham}
\end{equation}
with $\hat{\psi}_i^\dagger$ and $\hat{\psi}_i(i = a, b, m)$ the creation and annihilation operators. $\Delta$ is the intermediate-level detuning, $\phi_L$ is a relative laser-field phase and can be locked to a fixed value, {\it e.g.} $\phi_L=0$ by using phase locked laser beams \cite{Jones07}. Diagonalizing Hamiltonian (\ref{Ham}) allows us to obtain eigenvalues and eigenstates of a system, which can be described as one dark state with $E_d=0$ and $\left\vert d\right\rangle=\cos\theta \left\vert a\right\rangle-\sin\theta \left\vert b\right\rangle$, and two bright states with $E_{b_1}=(\Delta-\sqrt{\Delta^2+\Omega^2})/2$, $E_{b_2}=(\Delta+\sqrt{\Delta^2+\Omega^2})/2$ and $\left\vert b_1\right\rangle=\sin\theta\cos\gamma\left\vert a\right\rangle+\cos\theta\cos\gamma\left\vert b\right\rangle-\sin\gamma\left\vert m\right\rangle$, $\left\vert b_2\right\rangle=\sin\theta \sin\gamma\left\vert a\right\rangle+\cos\theta \sin\gamma\left\vert b\right\rangle+\cos\gamma\left\vert m\right\rangle$. The mixing dynamic angles are
\begin{equation}
\theta=\arctan[\frac{\Omega_1(t)}{\Omega_2(t)}],\gamma=\arctan[\frac{(\sqrt{\Delta^2+\Omega(t)^2}-\Delta)}{\Omega(t)}] 
\end{equation}
with $\Omega(t)=\sqrt{\Omega_1(t)^2+\Omega_2(t)^2}$. Without loss of generality, we will focus on $\left\vert b_1\right\rangle$ and $\Delta>0$ due to the symmetry of $E_{b1}$ and $E_{b2}$ with respect to $\Delta$.

The essence of double-STIRAP can be understood as follows. We initialize the system by preparing the atoms in a superposition state $\left\vert\Phi_{t=0}\right\rangle = \alpha\left\vert a\right\rangle + \beta\left\vert b\right\rangle$ where the population of state $\left\vert a\right\rangle$ ($\left\vert b\right\rangle$) is $P_{a(b)}^{t=0}=|\alpha|^2$($|\beta|^2$), giving to an unequal preparation $\Delta P_{ab}^{t=0}=P_a^{t=0}-P_b^{t=0}$. Here we assume $\Delta P_{ab}^{t=0}\geq0$. Such a superposition state can be yielded by rotation of single state through the ways of adiabatic passage \cite{Vitanov99,Vewinger03,Niu04,Kis05} or spontaneous radiation \cite{Chen15}. Atoms of {\it b}-state will  experience an ordinary d-STIRAP process which is irrelevant to the spontaneous loss from the intermediate state $\left\vert m\right\rangle$. This process is performed by a partially-overlapped laser pulse where $\Omega_1(t)$ precedes $\Omega_2(t)$, counter-intuitively with respect to {\it b}-state atoms. The resulting dynamical angle $\theta$ varies adiabatically from $\pi/2$ to 0 for $t\in [0,+\infty)$ and the system evolves along the dark state $\left\vert d\right\rangle$ yielding $\left\vert b\right\rangle\to\left\vert a\right\rangle$. In contrast, for {\it a}-state atoms this pulse sequence is intuitive, the system remains in bright state $\left\vert b_1\right\rangle$ while changing $\theta$. Under the prerequisite of the large intermediate detuning, it leads to a suppression of population in $\left\vert m\right\rangle$ because of $\gamma\to 0$. In this case the reduced eigenstate $\left\vert b_1\right\rangle\approx \sin\theta\left\vert a\right\rangle+\cos\theta\left\vert b\right\rangle$ acting as a dark state can realize an inversed population transfer $\left\vert a\right\rangle\to\left\vert b\right\rangle$ by b-STIRAP. 
Thus, the success of this double-STIRAP mainly lies on the use of a single pulse sequence with a large $\Delta$, although it is also the imperfection of double-STIRAP scheme that is largely affected by $\Delta$.

The steady-state population $|\phi_i^t|^2$ of bare state $\left\vert i\right\rangle$ versus the time $t$ are roughly plotted in Fig.\ref{modelfig}(b-c). For comparison, we use $\Delta=0.1\Omega$ (solid line) and $\Delta=10\Omega$ (dashed line). In d-STIRAP [(b)], the transfer efficiency from $\left\vert a\right\rangle$ to $\left\vert b\right\rangle$ is absolute, which is unaffected by the value of $\Delta$ due to $|\phi_m^t|^2=0$.  However, in (c) for b-STIRAP, due to $|\phi_m^t|^2=|sin\gamma|^2$ and $\gamma$ depends on $\Delta$, a small or zero detuning $\Delta$ will cause a large number of population occupied on $\left\vert m\right\rangle$, strongly breaking the coherence between two lower states. One way to solve the problem is to adopt a large detuning, {\it e.g.} $\Delta=10\Omega$ as shown in Fig. \ref{modelfig}(c), perfectly agreeing with the above theoretical predictions.

%To briefly summarize, our target is to switch populations of states $\left\vert a\right\rangle$ and $\left\vert b\right\rangle$ at a same time, it requires a d-STIRAP process for b-state atoms and a b-STIRAP process for a-state atoms that both STIRAPs work with a high performance. According to the analysis of eigenstates and eigenvalues, it suggests that $\Delta$ is a large and positive value.

\subsection{Adiabatic condition}

\begin{figure}
\includegraphics[width=2.8in,height=2.4in]{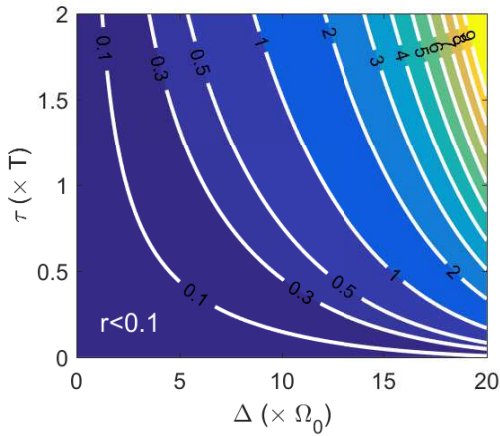}
\caption{(color online). Adiabatic parameter $r(t_p)$ at a critical time $t_p=(t_1+t_2)/2$ as a function of $\Delta$ and $\tau$. Here $\chi = \Omega_1^0/\Omega_2^0$, $\Omega_0 =\Omega_2^0 \sqrt{1+\chi^2}$ and $\Omega_0 T=100$. The region where $r<0.1$ is denoted.}
\label{adia}
\end{figure}

The existence of a proper steady state can not guarantee a perfect practical evolution towards the target state unless the adiabatic condition is also fulfilled. Accounting for the basic concept of adiabaticity \cite{Pu07} that requires the change of quantum state to be much slower than the energy difference between the target and its nearest eigenstates, a global form of adiabatic criterion is readily given by \cite{Grigoryan12}
\begin{equation}
|E_{b1}|\gg \max[ \left\vert\left\langle b1\right\vert\left\vert\dot{d}\right\rangle\right\vert,\left\vert\left\langle d\right\vert\left\vert\dot{b1}\right\rangle\right\vert ]
\end{equation}
resulting in a common analytical form 
\begin{equation}
r=\frac{\sqrt{2}\left\vert \dot{\Omega_{1}}\Omega_2-\Omega_1\dot{\Omega_2}\right\vert}{\Omega((\sqrt{\Delta^2+\Omega^2}-\Delta)^{3/2}(\Delta^2+\Omega^2)^{1/4})}\ll 1
\label{adian}
\end{equation}

To clearly represent the properties of adiabatic criterion (\ref{adian}), we adopt two Gaussian-type lasers  $\Omega_{1,2}=\Omega_{1,2}^{0}\exp[-(t-t_{1,2})^2/T^2]$ with $\Omega_{1,2}^0$ the peak intensity,  $t_{1,2}$ the central position and $T$ the pulse width. For pulse sequence that overlaps partly, it is assumed that the pulse delay $\tau=|t_1-t_2|$ and $\chi=\Omega_1^0/\Omega_2^0=1$. Figure \ref{adia} plots the dependence of adiabatic parameter $r$ on the parameters $\tau\in(0,T)$ and $\Delta\in(0,50\Omega_0)$ at a critical time $t_p=(t_1+t_2)/2$ where the interstate transition occurs. It is clearly shown that,  increasing the overlap region of pulses by reducing the pulse delay $\tau$ can improve the value $r(t)$, allowing $r\ll 1$ ($r<0.1$). If $\tau$ is optimally chosen and meanwhile the intermediate-level detuning $\Delta$ is large [e.g. $\Delta>10\Omega_0$], it can guarantee that the transfer process is not suffering from the loss from the excited state, realizing an efficient manipulation of superposition-state population. The importance of $\tau$ and $\Delta$ is clearly shown here.

\section{real dynamics and properties}

We now study the realistic performance of double-STIRAP by solving the master equation. For an arbitrary initial state $\left\vert \Psi_{t=0}\right\rangle =\alpha\left\vert a\right\rangle + \beta \left\vert b\right\rangle $ satisfying $\Delta P^{t=0}_{ab} >0$ and $|\alpha|^2+|\beta|^2=1$. An ideal reverse operation via double-STIRAP will lead to $\left\vert \Psi_{t=+\infty}\right\rangle =e^{i\delta\phi}\beta\left\vert a\right\rangle + \alpha \left\vert b\right\rangle $ with $\delta\phi$ a relative accumulated phase in the atom-field operation that are easily controllable in experiment \cite{Moller07}. \textbf{In the following we assume the change of phase is irrelevant and focus on exact population transfer of each state.} To characterize the practical population evolution, we introduce the switch fidelity $\kappa$ defined by
\begin{equation}
\kappa =\frac{\Delta P_{ba}^{t\to\infty}}{\Delta P_{ab}^{t=0}}= \frac{P_b^{t\to\infty}-P_a^{t\to\infty}}{P_a^{t=0}-P_b^{t=0}}
\label{kappa}
\end{equation}
with $P_i^t$ the population of state $\left\vert i\right\rangle$ at time $t$. The value $\kappa$ is expected to be in the range of $\kappa\in[-1,1]$ and if $\kappa \in(0,1]$ the switching process is pretty good and it tends to a perfect switch as $\kappa \to1$. On the other hand, if $\kappa \in[-1,0]$ the fidelity is poor and the limit case is no switch occurs when $\kappa = -1$. $P_i^t(=\rho_{ii}^t)$ is numerically solved from the master equation: $\dot\rho =-i [H,\rho]+L[\rho]$ where the Lindblad operator is $L[\rho] = \sum_{i=a,b}\Gamma_{mi}(\left\vert i\right\rangle \left\langle m\right\vert \rho \left\vert m\right\rangle \left\langle i\right\vert-\frac{{\left\vert m\right\rangle\left\langle m\right\vert},\rho}{2})$
and $\rho$ is the density matrix with the diagonal element $\rho_{ii}^{t}$ describing the population of $\left\vert i\right\rangle$ at time {\it t} conserved by $\rho_{aa}^{t}+\rho_{mm}^{t}+\rho_{bb}^{t}=1$. The spontaneous dissipation of $\left\vert m\right\rangle$ is denoted by the rates $\Gamma_{mi}$ for the transition of $\left\vert m\right\rangle\to\left\vert i\right\rangle$. The resulting dynamical equation for the density matrix is
\begin{eqnarray}
&i\dot{\rho}_{aa}&= \frac{\Omega_1}{2}(\rho_{ma}-\rho_{am})+i\Gamma_{ma}\rho_{mm} \\
&i\dot{\rho}_{bb}& = \frac{\Omega_2}{2}(\rho_{mb}-\rho_{bm})+i\Gamma_{mb}\rho_{mm}\\
&i\dot{\rho}_{am}&=\frac{\Omega_1}{2}(\rho_{mm}-\rho_{aa})-(\Delta+ i\frac{\Gamma_{ma}+\Gamma_{mb}}{2})\rho_{am}-\frac{\Omega_2}{2}\rho_{ab}\\
&i\dot{\rho}_{ab}&=\frac{\Omega_1}{2}\rho_{mb} - \frac{\Omega_2}{2}\rho_{am}\\
&i\dot{\rho}_{mb}&=\frac{\Omega_2}{2}(\rho_{bb}-\rho_{mm})+(\Delta- i\frac{\Gamma_{ma}+\Gamma_{mb}}{2})\rho_{mb}+\frac{\Omega_1}{2}\rho_{ab}
\end{eqnarray}
Solving Eqs.(6-10) numerically leads to the time-dependent population of each state.

\begin{figure}
\includegraphics[width=3.4in,height=3.8in]{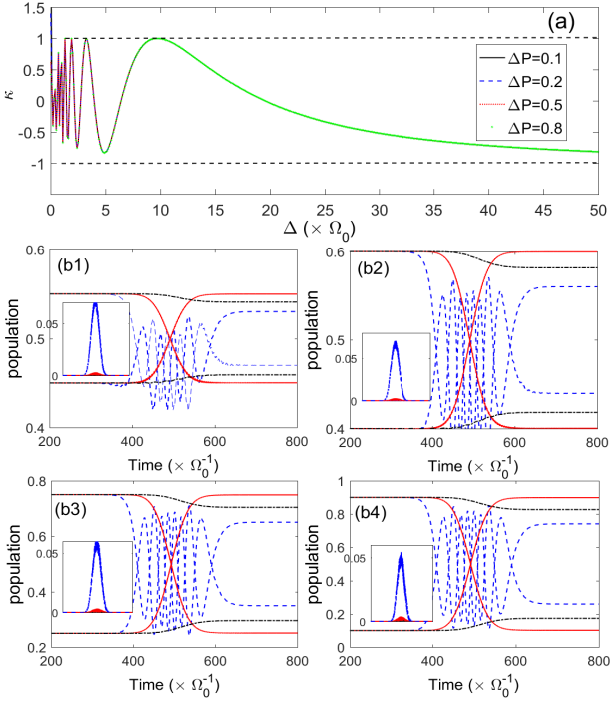}
\caption{(color online).  (a) Fidelity $\kappa$ as a function of $\Delta$ for different superposition-states $\Delta P_{ab}^{t=0}$; (b1-b4) Real population dynamics $P_{i}^t(i=a,m,b)$ for arbitrary initial preparations (b1) $\Delta P_{ab}^{t=0}=0.1$, (b2) $\Delta P_{ab}^{t=0} =0.2 $, (b3) $\Delta P_{ab}^{t=0}=0.5$, (b4) $\Delta P_{ab}^{t=0}=0.8$. The pulse delay is $\tau = 0.2T$, the intermediate-level detuning is $\Delta=\Omega_0$ (blue dashed line), $10\Omega_0$ (red solid line) and $50\Omega_0$ (black dash-dotted line). Insets: plots of $P_{m}^t$. The spontaneous loss of the excited state is ignored here.}
\label{dynamics}
\end{figure}

The sensitivity of fidelity $\kappa$ with respect to $\Delta$ is shown in Figure \ref{dynamics}(a) where the initial preparations $\Delta P_{ab}^{t=0}$ are arbitrary and found to be irrelevant. 
By varying the detuning $\Delta$, $\kappa$ exhibits a strong oscillation and sensitivity in the case of $\Delta<10\Omega_0$ owing to the influence from the intermediate excited state $\left\vert m\right\rangle$ although the adiabatic parameter $r$ is pretty good there. The oscillation frequency of $\kappa$ is shown to be proportional to $\Delta^{-1}$ as the effective Rabi frequency of a three-level system can be described by $\Omega_1\Omega_2/4\Delta$ \cite{Browaeys16}. For a large $\Delta$ where the adiabaticity is broken ($r>0.1$ as $\Delta>10\Omega_0$), the fidelity $\kappa$ decreases strongly with $\Delta$, reaching as low as -1 (no switch occurs). Owing to its sensitivity, an optimal control for the intermediate-state detuning is found to be significantly important in a real operation, also see Subsection \rm{IVB}.

To further identify the effectiveness of the adiabatic condition we comparably study the population $P_{i}^t$ of state $\left\vert i\right\rangle$ for three different $\Delta$ values and a fixed $\tau = 0.2T$ is used where the initial population is arbitrary in the simulation, here $\Delta P_{ab}^{t=0}$=$0.1,0.2,0.5,0.8$. 
In Figure \ref{dynamics}(b1-b4), for $\Delta = \Omega_0$ (blue dashed), although $r$ is quite good in this case, the effect of the excited state can not be neglected for $\Delta$ is small, which causes a strong oscillation before settling the steady state in the transition region, giving rise to a low fidelity $\kappa=0.602$. That is sensitive to the detuning $\Delta$ as displayed in Fig. \ref{dynamics}(a). For a large detuning $\Delta = 50\Omega_0$ (black dash-dotted) due to the breakup of adiabaticity accompanied by the elimination of the middle state, no clear switching is revealed and the final transfer fidelity is as low as $\kappa=-0.818$ ($\kappa<0$ means no switch occurs) although the condition of b-STIRAP is preserved. Interestingly, in between these two limits we study the case of $\Delta = 10\Omega_0$ (red solid) guaranteeing the requirements of both double-STIRAP transition and adiabaticity. It is obvious that a perfect switch for the superposition-state population is robustly established with a high-fidelity of almost 1.0 ($\kappa = 0.996$), which is exactly irrelevant to the initial preparations. 

\textbf{Similar results were demonstrated in Ref. \cite{Du14} where authors theoretically and experimentally realized a reverse operation for a superposition state $\Delta P_{ab}^{t=0}=0.5$ with near-unity fidelity when the required adiabatic condition is fulfilled. Their theoretical study showed a strong oscillation during the intermediate region due to the phase difference between $\left\vert d \right\rangle$ and $\left\vert b_1 \right\rangle$ while fast oscillations were not clearly observed in practical experiment. In the present work, for $\Delta=10\Omega_0$ (red solid) which meets the requirements for both adiabaticity and double-STRAP transition, we show the switching process is perfect with no oscillations. The effect of phase difference will be discussed in end of Section \rm{V}.}

\section{Parameter constraints}

We further explore the robustness of the scheme against the variation of some parameters that is controllable in experiments. To this end, we theoretically calculate the fidelity of the switch under the change of the ratio between spontaneous losses $\alpha$, the ratio between Rabi frequencies $\chi$, the intermediate-level detuning $\Delta$ as well as the delay between pulse sequence $\tau$, while keeping other parameters unchanged.

\subsection{Effect of ratios $\alpha$, $\chi$ on fidelity}

\begin{figure}
\includegraphics[width=3.4in,height=1.6in]{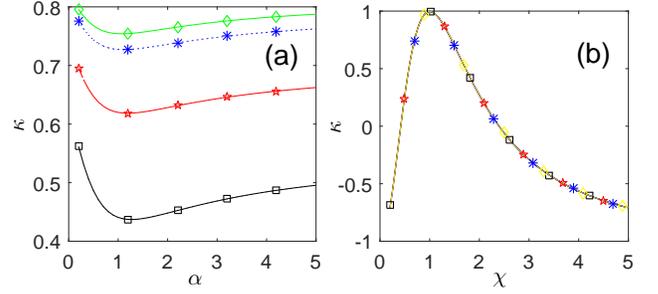}
\caption{(color online). (a) Fidelity $\kappa$ versus the variation of relative spontaneous decay rate: $\alpha= \Gamma_{ma}/\Gamma_{mb}$ ($\Gamma_{mb} = \Omega_0$). From top to bottom, different initial preparations are considered, which are $\Delta P_{ab}^{t=0} = 0.8$ (green dash-dotted line with diamonds), 0.5 (blue dotted line with stars), 0.2 (red dashed line with stars), 0.1(black solid line with squares). (b) Fidelity $\kappa$ versus the relative amplitude of two lasers: $\chi=\Omega_{1}^0/\Omega_{2}^0$. Other parameters are same as (a).}
\label{decay}
\end{figure}

The fidelity $\kappa$ as a function of the relative spontaneous loss $\alpha = \Gamma_{ma}/\Gamma_{mb}$ for different initial preparations is displayed in Fig.\ref{decay}(a) while keeping $\Gamma_0 =\sqrt{\Gamma_{ma}^2+\Gamma_{mb}^2}= 1.0\Omega_0$ and $\alpha\in(0.2,5.0)$. Fig.\ref{decay}(a) shows that $\kappa$ is very sensitive to the unequal initial preparations, that is a big difference of population will yield a larger $\kappa$, i.e. $\kappa(\Delta P_{ab}^{t=0}=0.8)>\kappa(\Delta P_{ab}^{t=0}=0.5)>\kappa(\Delta P_{ab}^{t=0}=0.2)>\kappa(\Delta P_{ab}^{t=0}=0.1)$. To understand this, if $\Delta P_{ab}^{t=0}$ is large, it means state $\left\vert a\right\rangle$ is dominately prepared, allowing {\it a}-state atoms largely converted via the process of single b-STIRAP, rather than by double-STIRAP. When a small $\Delta P_{ab}^{t=0}$ is used for comparable population in both states $\left\vert a\right\rangle$ and $\left\vert b\right\rangle$, the double-STIRAP scheme plays a significant role which is greatly affected by the decays from the intermediate state, giving rise to a relative lower $\kappa$. For same $\Delta P_{ab}^{t=0}$, $\kappa$ is also slightly varying with the ratio $\alpha$. It is shown that, $\kappa$ at $ \Gamma_{ma}\approx\Gamma_{mb}$ ($\alpha =1$) is smaller than at $ \Gamma_{ma}\neq\Gamma_{mb}$ ($\alpha\neq1$), mainly caused by the competition between two different STIRAP transitions.

For comparison, $\kappa$ versus the variation of relative laser amplitudes $\chi$ for different initial conditions is illustrated in Fig. \ref{decay}(b) where $\Omega_2^0 = \Omega_0$ and the decays are ignored. Greatly differing from (a), here $\kappa$ reveals a perfect irrelevance to the initial condition because two ground-state atoms $P_{a}^{t=0}$, $P_{b}^{t=0}$ are transfered individually by single d- or b-STIRAP although a common pulse sequence $\Omega_1(t)$ and $\Omega_2(t)$ is shared by them. To this end, $\kappa$ exhibits a high symmetry with respect to $\chi=1$ while arbitrarily adjusting the amplitudes of two lasers, {\it e.g.} $\kappa(\chi=0.2) = \kappa(\chi = 5)$. The maximum of $\kappa$ exists if $\chi=1$ presenting the significance of using same peak amplitude of the laser pulses, with the biggest overlap region for improving the switch fidelity.

\subsection{Optimal control for $\Delta$ and $\tau$}

\begin{figure}
\includegraphics[width=2.8in,height=2.4in]{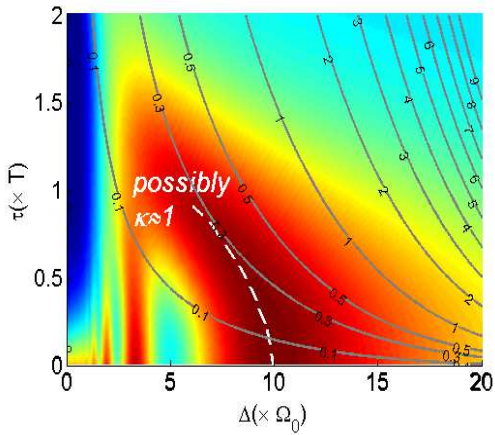}
\caption{(color online). Fidelity $\kappa$ versus the variations of $\Delta$ and $\tau$. For a given $\tau$, the maximal $\Delta_{max}$ required for attaining the local maximum $\kappa\approx 1.0$ is shown by a white dashed curve. The spontaneous loss is $\Gamma_{ma} = \Gamma_{mb} = 0.1\Omega_0$ and an unequal population is $\Delta P_{ab}^{t=0} = 0.5$. For comparison, curves denoted by gray colors repeatedly plot the results of adiabatic parameters $r$ as displayed in Fig. \ref{adia}. }
\label{adiady}
\end{figure}

In section \rm{IIB} we have studied the dependence of adiabaticity $r$ on $\Delta$ and $\tau$, presenting the importance of $\Delta$ and $\tau$ that must meet the conditions of b-STIRAP and adiabaticity at the same time. In Fig.\ref{dynamics}(a) a strong sensitivity (dynamical oscillation with same amplitude due to $\Gamma_{ma(b)}\approx 0 $) of $\kappa$ with respect to the change of $\Delta$ has been displayed and we pay attention to the maximal permission of $\Delta_{max}$ for a given $\tau$, where $\kappa$ attains its local maximal. From Fig.\ref{dynamics}(a) it implies that the maximal switch fidelity $\kappa$ may be obtained at $\Delta_{max}=10\Omega_0$ and $\tau = 0.2T$ when the decays are ignored.

By continuously varying $\Delta$ and $\tau$, we numerically calculate the fidelity $\kappa$ in Fig. \ref{adiady} where the position of the local maximal $\kappa$ for given $\Delta_{max}$ and $\tau$ is plotted by a white dashed curve. 
For comparison, the adiabaticity contour as same as Fig. \ref{adia} is also plotted by gray curves. When the pulse delay $\tau$ is small, a strong oscillation of $\kappa$ with rapidly decaying amplitude as $\Delta$ decreases is shown, stressing the importance of optimal control for the detuning $\Delta$, to achieve a perfect switching. As increasing $\tau$, we see the oscillation of $\kappa$ disappears however it brings $\kappa\to -1$ (dark blue region) once $\Delta$ is very small. The reason for that is mainly caused by a big loss from the intermediate state in the one-photon resonance case, although the adiabaticity there is quite good $r<0.1$.

\textbf{Moreover, by comparing $\kappa$ and $r$ in Fig. \ref{adiady}, we notice that, for realizing a high fidelity $\kappa\approx 1$, adiabaticity is a very important criterion since the boundary for local maximal value of $\kappa$ requires the condition of $r<0.3$ to keep a quite good adiabaticity. However, this boundary is only partly agrees with the contour of $r=0.3$, mainly because for the case of small detuning $\Delta$ (keeping $r<0.3$), a near-resonant one-photon excitation with respect to $\left\vert m\right\rangle$ will cause a big loss of population that leads to a poor switching efficiency. The results of Fig. \ref{adiady} imply that to realize a perfect switch, the conditions of adiabaticity and a suitable large intermediate detuning (i.e. the b-STIRAP transition) are required.}

.

\section{scheme Implementation}

\begin{figure}
\includegraphics[width=2.8in,height=2.3in]{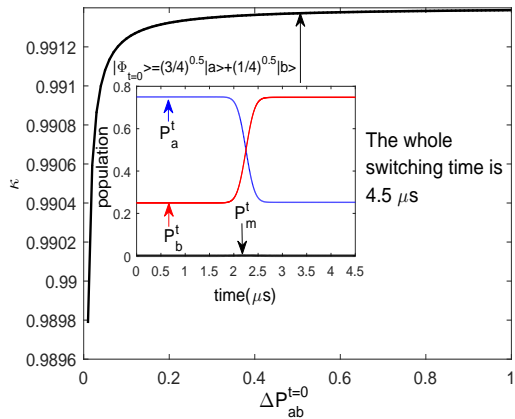}
\caption{Simulation for the fidelity $\kappa$ as a function of $\Delta P_{ab}^{t=0}$ in a real experimental implementation. Inset: a specific example for the population dynamics with the initial state $\left\vert\Phi_{t=0} \right\rangle=\frac{1}{2}\left\vert a\right\rangle+\frac{\sqrt{3}}{2} \left\vert b\right\rangle$. The whole operation time is only 4.5 $\mu$s. All parameters are described in the text.}
\label{exp}
\end{figure}

Finally we turn to the discussion of the current experimental feasibility. We consider the implementation to arbitrary superposition-state population prepared by coherent Rabi oscillation or one-qubit rotation. Simulating experimental results in the manipulation of the reverse operation for the superposition-state switching is important. For $^{87}$Rb, the atomic states that could be used for superposition-state preparations are the F=1 ($=\left\vert a\right\rangle$) and F=2 ($=\left\vert b\right\rangle$) hyperfine levels of the 5S$_{1/2}$ ground state. The intermediate state is an excited state F=0 ($=\left\vert m\right\rangle$) of 5P$_{3/2}$ \cite{Ficek04}, and its the spontaneous decays is typically $\Gamma_{ma}=\Gamma_{mb}=6.1$MHz. In the simulation, the amplitude of the laser pulse sequence is $\Omega_1^0 = \Omega_2^0  = 2\pi\times 25$MHz, the pulse width is $T=0.45\mu$s, the delay of pulse sequence is $\tau = 0.9$ns, giving rise to the central positions of lasers $t_2 = 2.25\mu$s, $t_1=2.2491\mu$s. The intermediate-level detuning is $\Delta = 2\pi\times353$MHz accessible by most current experiments, {\it e.g.} \cite{Ciamei17}.

As shown in Fig.\ref{exp}, the fidelity $\kappa$ in a real operation at time $t=4.5\mu$s (the end of pulse sequence) as a function of the initial unequal population $\Delta P_{ab}^{t=0}\in(0,1.0)$ is represented. It is remarkable that $\kappa>0.99$ (near-perfect fidelity) is preserved for arbitrary preparations, suggesting the robustness of our scheme in the applications of coherent population switching and manipulation. We also show a realistic time-dependent population switching under experimental parameters for a determined superposition-state $\left\vert\Phi_{t=0} \right\rangle=\frac{1}{2}\left\vert a\right\rangle+\frac{\sqrt{3}}{2} \left\vert b\right\rangle$ ($\Delta P_{ab}^{t=0}=0.5$). A perfect switching for the initial populations is revealed with $P_m^t\approx 0$ for the complete evolution. Note that the whole switching time required here is only 4.5 $\mu$s, much smaller than that reported in Du's experiment ($T=2ms$) \cite{Du14}.

\textbf{Before ending, we will briefly discuss the influence of phase in the switching of superposition state. If the relative laser-field phase $\phi_L$ is time-dependent, the resulting accumulated relative phase $\delta\phi$ is nonvanishing. As a result by setting $\delta\phi=0$ or $\delta\phi=\pi$, the initial superposition state $\left\vert\Phi_{t=0} \right\rangle=\frac{1}{2}\left\vert a\right\rangle+\frac{\sqrt{3}}{2} \left\vert b\right\rangle$ will evolve into the symmetric superposition state $\left\vert\Phi_{t=\infty} \right\rangle=\frac{\sqrt{3}}{2}\left\vert a\right\rangle+\frac{1}{2} \left\vert b\right\rangle$ or the antisymmetric superposition state $\left\vert\Phi_{t=\infty} \right\rangle=\frac{\sqrt{3}}{2}\left\vert a\right\rangle-\frac{1}{2} \left\vert b\right\rangle$ by adiabatically following the double-STIRAP transition. Furthermore, the geometric (berry) phases acquired during the evolution can be varied by the laser fields, offering special uses for phase gate implementation, which will be left for a next-step study.}

\section{Conclusion}

We propose a simple and practical method to fast and efficiently switch the population of a superposition state consisted of two ground states via a novel double-STIRAP protocol. Differing from previous double STIRAP systems that we only utilize a single pulse sequence that plays the role of both d-STIRAP and b-STIRAP pulses. The key for implementing this double-STIRAP is the optimal control for the intermediate-level detuning $\Delta$ and the delay of pulse sequence $\tau$, allowing to satisfy the b-STIRAP transition and adiabaticity at the same time. The robustness of the scheme is verified by studying the dependence of the switch fidelity with respect to the ratio between spontaneous decays of excited state, as well as the ratio between Rabi frequencies, which relaxes the requirement for specific choice of relevant parameters in a realistic operation, as long as $\Delta$ and $\tau$ are well-determined. Finally we show that, under the experimental parameters, it is feasible to achieve a robust quantum superposition-state population switching with near-unit fidelity in a total evolution time $t=4.5\mu$s for arbitrary initial populations, which may offer interesting perspectives as a basic building block for quantum state control in the field of quantum information processing.

\section{Acknowledgements}

This work is supported by the NSFC under Grants No. 11474094, No. 11104076, and No. 11574086,
by the Science and Technology Commission of Shanghai Municipality under Grant No. 18ZR1412800 and No. 16QA1401600, 
the Specialized Research Fund for the Doctoral Program of Higher Education No. 20110076120004.

%\end{multicols}{2}
\end{document}